# Bismuth ferrite as low-loss switchable material for plasmonic waveguide modulator

Viktoriia E. Babicheva[1,2,*], Sergei V. Zhukovsky[1,2], and Andrei V. Lavrinenko[1]

[1]*DTU Fotonik, Technical University of Denmark, Oersteds Plads 343, 2800 Kgs. Lyngby, Denmark*

[2] *ITMO University, Kronverkskiy, 49, St. Petersburg 197101, Russia*

[*] E-mail: baviev@gmail.com

**Abstract**. We propose new designs of plasmonic modulators, which can be utilized for dynamic signal switching in photonic integrated circuits. We study performance of plasmonic waveguide modulator with bismuth ferrite as an active material. The bismuth ferrite core is sandwiched between metal plates (metal-insulator-metal configuration), which also serve as electrodes so that the core changes its refractive index under applied voltage by means of partial in-plane to out-of-plane reorientation of ferroelectric domains in bismuth ferrite. This domain switch results in changing of propagation constant and absorption coefficient, and thus either phase or amplitude control can be implemented. Efficient modulation performance is achieved because of high field confinement between the metal layers, as well as the existence of mode cut-offs for particular values of the core thickness, making it possible to control the signal with superior modulation depth. For the phase control scheme, $\pi$ phase shift is provided by 0.8-μm length device having propagation losses 0.29 dB/μm. For the amplitude control, we predict up to 38 dB/μm extinction ratio with 1.2 dB/μm propagation loss. In contrast to previously proposed active materials, bismuth ferrite has nearly zero material losses, so bismuth ferrite based modulators do not bring about additional decay of the propagating signal.

## 1. Introduction

Plasmonic structures were shown to provide advantages for waveguiding and enhanced light-matter interaction, as utilizing surface plasmon waves at metal-dielectric interface allows efficient manipulation of light on the subwavelength scale [1-4]. A metal-insulator-metal (MIM) waveguide provides the most compact configuration due to high mode localization within the dielectric core, and consequently efficient interaction between field of the mode and active material if it is placed between metal layers [5-7]. Although detailed characterization of the devices encounters issues because of the small mode size and high insertion losses, it has been shown recently that the efficient coupling from a photonic waveguide to an MIM structure can be realized to launch the signal [8]. Moreover, different approaches have been proposed, for example confining light either by thick metal layers or by more specifically designed metamaterials, for instance hyperbolic metamaterials [9,10].

Plasmonic waveguide modulators and switches are of major interest for ultra-compact photonic integrated circuits and have been widely studied last several years [11,12]. Several promising designs have been proposed including investigation of various active materials, such as silicon [13-16], transparent conductive oxides (TCOs) [7,17-21], graphene [22], nonlinear polymers [23], thermo-optic polymers [24,25], gallium nitride [26,27], and vanadium dioxide [28-32]. Some of them were shown to outperform conventional photonic-waveguide-based designs in terms of compactness [19,33].

In general, one can distinguish two classes of active materials according to physical mechanisms underlying in refractive index control: carrier concentration change (e.g. TCOs, silicon, and graphene) and nanoscale structural transformations (e.g. gallium and vanadium dioxide). For example, TCOs provide a large change of refractive indexes and can be utilized for fast signal modulation on the order of several terahertz [7,17-21,34-36]. However, they possess high losses, consequently the modal propagation length is fairly small [17,18]. For loss mitigation, one can implement gain materials and directly control the absorption coefficient [37,38], but such active materials can significantly increase the noise level.

In contrast to carrier concentration change, structural transformations cannot provide such high bit rate, and megahertz operation is expected due to microsecond timescales of the transformations. Yet, refractive index changes that accompany nanoscale material transformations are much higher than those caused by carrier concentration change. In particular, extinction ratio up to 2.4 dB/μm was demonstrated for a hybrid plasmonic modulator based on metal – insulator phase transition in vanadium dioxide [31].

Ferroelectric materials, such as bismuth ferrite ($BiFeO_3$, BFO) or barium titanate ($BaTiO_3$, BTO), possess promising features for optical modulation [39-47]. Under applied voltage, the ferroelectric domains can be partially reoriented from the in-plane orientation (with an ordinary refractive index $n_o$) to the out-of-plane orientation (with extraordinary index $n_e$) [48,49]. Thus, the refractive index for a field polarized along one axis can be changed, and control of propagating signal is achieved. Variation of the applied voltage provides a varying degree of domain switching, and thus the required level of propagating signal modulation can be realized. BTO was shown to provide high performance for photonic thin film modulators [39-42], as well as electro-optic properties in plasmonic interferometer-based [43] and waveguide-based [45] modulators. However, BFO has higher birefringence with refractive index difference $\Delta n = 0.18$ nearly three times higher than in BTO. Recently a strong change of refractive index in BFO was demonstrated [44] and proposed for electro-optic modulation [46-47].

Here for the first time, we propose an implementation of BFO as the active materials for plasmonic waveguide modulators. We analyze different modulator designs based on MIM waveguide, and compare the performance of these modulators. Because of the low losses of BFO at telecom wavelength, one can achieve large phase shifts and high extinction ratio on a sub-micron length. Specifically, we predict a $\pi$ phase shift in low-loss phase modulator only 0.8 μm in length, and up to 38 dB/μm extinction ratio in a high-contrast absorption modulator. In Section 2, we analyze dispersion properties of an MIM waveguide with BFO core. In Section 3, phase and absorption modulation is studied in more detail. Section 4 follows with summary and conclusions.

## 2. Eigenmodes of the waveguide with BFO core

Schematic view of the MIM waveguide with BFO is shown in Fig. 1a. We are interested in modulation at the telecom wavelength, $\lambda_0 = 1.55$ μm, so the metal permittivity is fixed at $\varepsilon_m = -128.7 + 3.44i$ (silver, [50]). Plasmonic modes are defined by the equation [51]:

$$\tanh(qd) = -\left(\frac{k_m}{q}\frac{\varepsilon_{zz}}{\varepsilon_m}\right)^{\pm 1} \qquad (1)$$

where "$\pm$" corresponds to symmetric and antisymmetric modes, respectively; $d$ is the core thickness; $\varepsilon_{xx}$ and $\varepsilon_{zz}$ are components of the permittivity tensor of the core; $q = \sqrt{(\varepsilon_{zz}/\varepsilon_{xx})\beta_0^2 - \varepsilon_{zz}k_0^2}$; $k_m = \sqrt{\beta_0^2 - \varepsilon_m k_0^2}$; $k_0 = 2\pi/\lambda_0$ is the wave number in vacuum; $\beta_0 = \beta + i\alpha$ is the complex propagation constant to be determined. We consider $n_o = 2.83$ and $n_e = 2.65$ [44] and solve the dispersion equation of a three-layer structure. We consider two options for the device off-state: with $\varepsilon_{xx} = n_o^2$ and $\varepsilon_{zz} = n_e^2$ ("x-ordinary", denoted "ox") and with $\varepsilon_{xx} = n_e^2$ and $\varepsilon_{zz} = n_o^2$ ("z-ordinary", denoted "oz"); we assume that the imaginary part of $\varepsilon_{ij}$ is very small at 1.55 μm. Under applied voltage, the domains are reoriented, and both off-states switch to the same on-state (labeled "e") with tensor components equal to $\varepsilon_{xx} = \varepsilon_{zz} = n_e^2$ (Fig. 1b).

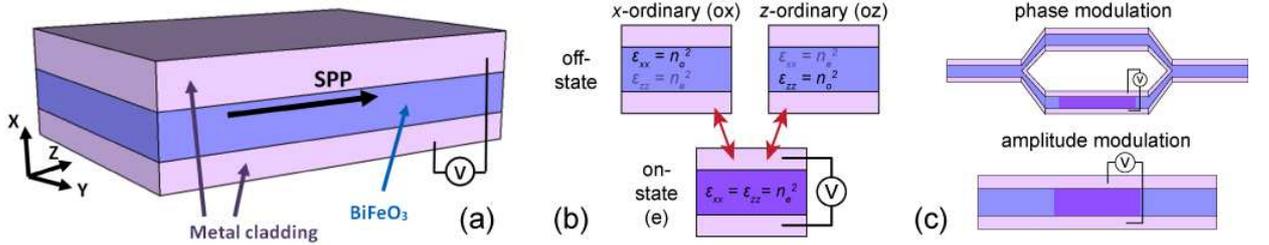

Fig. 1. (a) Schematic view of plasmonic modulator based on metal-insulator-metal waveguide with BFO core as active material. (b) Illustration of BFO switching in the "x-ordinary" and "z-ordinary" scenario. (c) Schematics of BFO-based plasmonic switches based on phase or absorption modulation principles.

We solve the dispersion equation (1) numerically for different core thickness $d = 50…400$ nm. As seen in Fig. 2, the structure supports three modes in the considered parameter range: two symmetric (denoted "$s_1$" and "$s_2$") and one antisymmetric (denoted "as"). The results show a significant change of propagation constant $\beta$ and absorption coefficient $\alpha$ for all the three modes during switching between either of the two off-states [(ox) and (oz)] and the on-state (e), when the refractive index along one of the axes changes from $n_o$ to $n_e$.

Such difference in the mode indexes can allow efficient operation of the device. We see that the mode "$s_1$" provides the maximum change of $\beta$ between the off-state "ox" and the on-state (Fig. 2a), as well as the lowest losses nearly uniform across the range of core thickness variation (Fig. 2b). Thus, this configuration is particularly favourable for a phase modulator in an interferometer-type setup (see Fig. 1c, top). On the other hand, the two remaining modes ("$s_2$" and "as") feature an abrupt step in the dependence $\alpha(d)$ near the cut-off values of the core thickness, occurring at different $d$ for on- vs. off-state (Fig. 2b). These modes are therefore particularly suitable for direct amplitude modulation in a waveguide-type device (Fig. 1c, bottom), especially when the off-state "oz" is used. In the following section, we perform a more detailed analysis of these regimes.

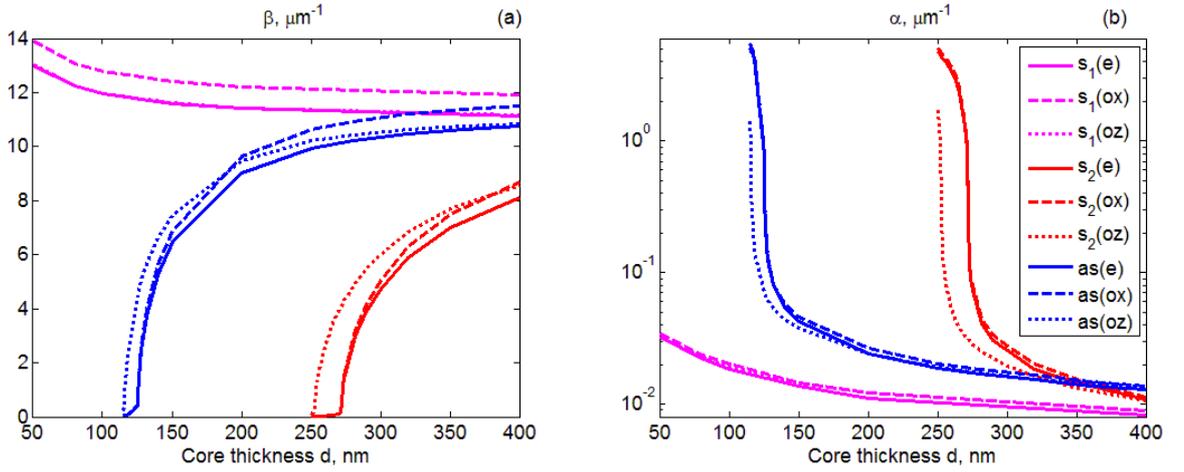

Fig. 2. Propagation constants (a) and absorption coefficients (b) for different modes of MIM waveguide: two symmetric ($s_1$ and $s_2$) and antisymmetric (as). Notations (ox), (oz), and (e) correspond to two off-states (x-ordinary and z-ordinary) and the on-state, respectively (see Fig. 1b). Legend is the same on both plots.

## 3. Modulator designs and performance characterization

### 3.1. Phase-modulation operation

The first symmetric mode "$s_1$" possesses the highest β, which varies in a range 11…14 μm$^{-1}$ and corresponds to the effective index $n_{eff}$ = 2.7…3.5. Absorption coefficient of this mode is the smallest and has almost no difference between the off- and on-states. Thus this mode is suitable for signal phase control (Fig. 1c, top)

We calculated the length required to achieve π phase shift $L_\pi = \pi/\beta_o - \beta_e$. It shows value around 4 μm in a broad range of $d$ and decreases with the decrease of core thickness $d$ (Fig. 3a). The propagation losses are 0.08…0.3 dB/μm and thus a short length remains relatively high transmission ($T_{dB} = 10lg(T_0/T) = 8.68\alpha L_\pi$, Fig. 3b).

Thus, the phase control can be put into practice via the Mach-Zehnder interferometer with the device length down to 4 μm. The operation bandwidth is large since the effect of mode index change is essentially non-resonant (near-flat lines in Fig. 3a) and since the BFO refractive index only slightly varies with the wavelength [44].

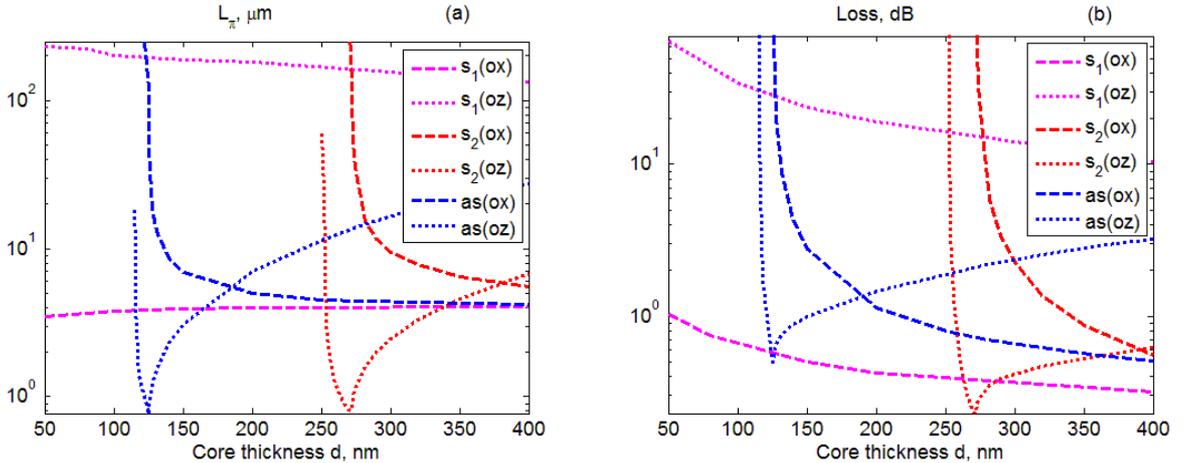

Fig. 3. (a) Device length $L_\pi$ needed to achieve π phase difference between on- and off-state. (b) Loss in the device with length $L_\pi$. The labels (ox) and (oz) correspond to the switching scenarios from the x- and z-ordinary off-state to the on-state (see Fig. 1b).

On the other hand, when the core thickness varies, the second symmetric "$s_2$" and antisymmetric "as" modes possess a much more abrupt change. They have very low β and high α for some particular thicknesses, which corresponds to the modes exhibiting cut-off (Fig. 2). We can define the mode as propagating when it satisfies the condition Q = β/α > 1. Thus "$s_2$" has cut-off at: $d(s_2,e)$ = 272 nm, $d(s_2,ox)$ = 272 nm, and $d(s_2,oz)$ = 253 nm. Because of the fast pronounced change near the cut-off thicknesses, the range $d$ = 253…272 nm corresponds to the largest difference between $\beta(s_2,e)$ and $\beta(s_2,oz)$. The length required for π phase shift is down to 800 nm (see Fig. 3a) and the mode losses are even lower than for "$s_1$" (see Fig. 3b). Similar properties are shown by mode "as" in a range of thicknesses 116…125 nm. Propagation losses of these two modes are 0.29 and 0.6 dB/μm. Thus, adopting BFO a low-loss ultra-compact plasmonic modulator can be realized.

*3.2. Absorption-modulation operation*

Another way to implement a plasmonic switch is the direct manipulation of absorption coefficient α. Because of the mode cut-off discussed previously, both modes "$s_2$" and "as" provide such a possibility (see Fig. 2b). One can define the figure of merit (FoM) of the device as $\text{FoM} = (\alpha_e - \alpha_o)/\alpha_o$. Both modes have high FoMs in the cut-off region, where there is an abrupt α change (Fig. 4a).

We see that for the symmetric mode "$s_2$", the FoM value of 67 is reached. At this point, the extinction ratio ER = $\alpha_e - \alpha_o$ = 20 dB/μm, and consequently 3 dB switch can be realized on 150 nm. Corresponding propagation length z = $1/(2\alpha_o)$ is 12 μm (Fig. 4b). For the asymmetric mode "as", the device characteristics are also prominent: FoM up to 29, extinction ratio of 28 dB/μm (allowing a 3 dB switch on 107 nm), and propagation length z = 3.5 μm (Fig. 4b).

For all the considered designs, modes are strongly localized within the core and efficient coupling from photonic waveguide [8] can be realized to launch symmetric mode. In case of special requirements from the design, launching of antisymmetric mode can be accomplished by coupling of another plasmonic waveguide with asymmetric mode.

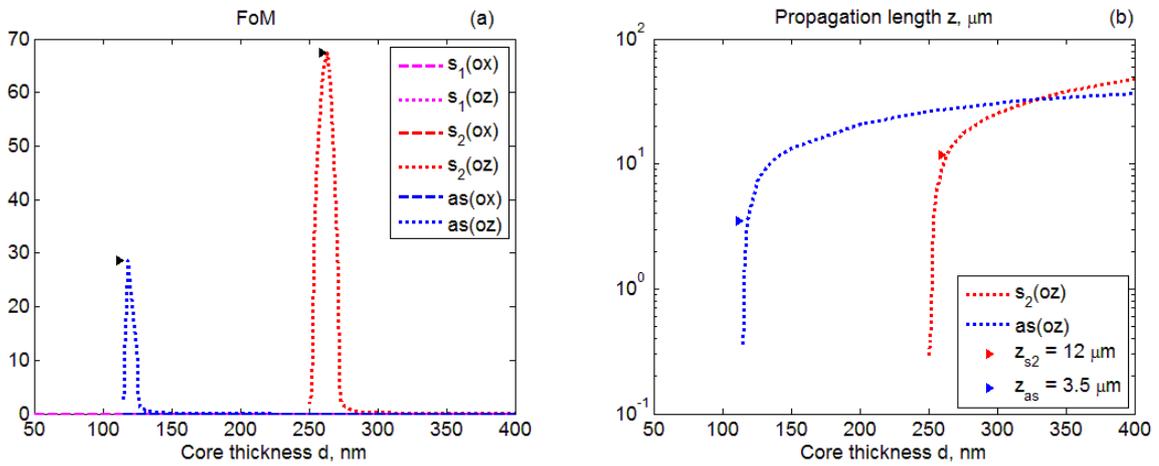

Fig. 4. (a) FoM and (b) propagation length z for the second symmetric "$s_2$" and asymmetric "as" modes. Triangular marks on (b) correspond to maximum FoM on (a).

## 4. Conclusion

In summary, we studied properties of the MIM plasmonic waveguide with the BFO core aiming to utilize such a waveguide as a building block for an efficient plasmonic modulator. The proposed designs give the following three advantages.

i) From the material point of view, low losses of BFO (nearly zero for λ > 1400 nm [44]) do not cause additional attenuation from waveguide core, and thus do not increase insertion loss of the whole device, in contrast to TCO or vanadium dioxide.

ii) From the design point of view, MIM configuration allows cut-off of the propagating mode and thus makes it possible to modulate propagation signal by switching it on and off through metal layers serving as electrodes.

iii) MIM configuration provides high confinement of the mode, and the field does not expand outside the waveguide.

The device can be realized in two ways (Fig. 1c). On the one hand, signal phase control can be implemented in Mach-Zehnder interferometer and device length 0.8…4 μm is required. Such a compact structure indicates the high potential of BFO-based devices. Operation with the first symmetric waveguide mode can be broadband as BFO refractive index is only slightly varying at wavelengths close to the telecom range, and mode characteristics do not have pronounced change when the core thickness varies. On the other hand, effective signal amplitude control by direct change of absorption can be realized. In this case a BFO plasmonic modulator can have FoM up to 67, which is comparable with recently reported values for modulators based on dielectric-loaded plasmonic waveguide with graphene [22], and is much higher than FoMs of previously reported devices based on indium tin oxide, vanadium dioxide, or InGaAsP active layers [18,29,37,38].

**Acknowledgments**

V.E.B. acknowledges financial support from SPIE Optics and Photonics Education Scholarship and Kaj og Hermilla Ostenfeld foundation. S.V.Z. acknowledges partial financial support from the People Programme (Marie Curie Actions) of the European Union's 7th Framework Programme FP7-PEOPLE-2011-IIF under REA grant agreement No. 302009 (Project HyPHONE).